\documentclass[fleqn,twoside]{article}
\usepackage{espcrc2,psfig}

\usepackage{graphicx}
\usepackage[figuresright]{rotating}

\newcommand{\AmS}{{\protect\the\textfont2
  A\kern-.1667em\lower.5ex\hbox{M}\kern-.125emS}}

\title{Exploring N=1 SYM(2+1): the Stress-Tensor Correlator}

\author{U.~Trittmann
	\thanks{Work done in collaboration with S.~Pinsky and J.~Hiller}\\
	Department of Physics, Ohio State University, 
	174 W 18th Ave, Columbus, OH, USA
        }

\begin{document}

\begin{abstract}
The evaluation of field theoretic correlators at strong couplings
is especially interesting in the light of recently discovered 
string/field theory correspondences.
We present a calculation of the stress-tensor correlator 
in ${\cal N}=1$ SYM theory in 2+1 dimensions. 
We calculate this object numerically with the method of supersymmetric 
discrete light-cone quantization (SDLCQ) at large $N_c$.
For small distances we reproduce the conformal field theory result
with the correlator behaving like $1/r^6$.
In the large $r$ limit 
the correlator is determined by the (massless) BPS states of the theory. 
We find a critical value of the coupling where 
the correlator goes to zero in this limit.
This critical coupling is shown to grow linearly with the square 
root of the transverse momentum resolution. 
\vspace{1pc}
\end{abstract}

\maketitle

\section{Introduction}

In the present note we will report on the evaluation of the 
correlator of the stress-energy tensor in three-dimensional 
${\cal N}=1$
supersymmetric Yang-Mills theory (SYM(2+1)).
Correlation functions play a crucial role in
string/field theory correspondences, because they can typically be 
calculated both on the string and on the field theory side. 
The crucial difficulty is that due to the nature of the correspondence,
the correlator has to be evaluated in the strong coupling regime
on one of the sides. In the past, we have used the non-perturbative method of 
supersymmetric discrete light-cone quantization (SDLCQ) to evaluate 
correlators at strong couplings, with encouraging results for the 
validity of the conjectured correspondence \cite{hlpt00}. 

While ${\cal N}=1$ SYM(2+1) is interesting by itself for a number of reasons,
it is the version of this theory with extended ${\cal N}=8$ supersymmetry
that corresponds to a string theory of D2 branes \cite{Itzhaki}.
We have learned from past calculations that the number of supersymmetry 
operators greatly enhances the size of the numerical calculations.
As a first step towards the full ${\cal N}=8$ calculations we 
tackle ${\cal N}=1$ SYM(2+1) here.

Contrary to the mass spectrum, 
correlators use all spectral information, namely energy eigenvalues and the 
eigenfunctions, and can thus be used to 
test wavefunctions. In the present case, this is especially useful in order
to examine the BPS states of the theory. These states are annihilated by
one of the supercharges, and their masses are protected by their symmetry
properties,
whereas their wavefunctions could change as a function of the coupling.
This is what we seem to find in the present study.

From conformal field theory it is known that the behavior of the correlator 
as a function of the distance $r$ between the operators should 
be like $1/r^6$ for small distances.
On the other hand, the contributions of 
bound states will have a characteristic length scale associated with their
size and one would therefore expect a $1/r^5$ behavior at small $r$.
The different contributions have thus to work in concert to give the 
correct $1/r^6$ behavior. We will reproduce this behavior both  
analytically by calculating the free particle correlator and numerically
by evaluating the correlator in the interacting theory.
At large distances we find a critical coupling where the correlator 
goes to zero.
In general we find a good convergence in both the longitudinal
and the transverse cutoffs, which 
is non-trivial since we are dealing here with a three-dimensional theory,
whereas most of our previous studies were in 1+1 dimensions.
From conformal field theory calculations one knows that the 
correlators are simpler in collinear limit $x_\perp\rightarrow 0$,
and we will work in this limit.


\section{SDLCQ of N=1 SYM(2+1)}

Discretized light-cone quantization (DLCQ) can be formulated in a way 
that preserves supersymmetry at each step of 
calculation \cite{Sakai95}. Namely, one discretizes the supercharge $Q^-$ 
rather than the Hamiltonian $P^-$.
This special approach is called supersymmetric DLCQ, or SDLCQ.
One uses light-cone coordinates
\begin{eqnarray*}
x^\pm\equiv\frac{1}{\sqrt{2}}(x^0\pm x^1), \qquad x^{\perp}&=& x^{\perp},
\end{eqnarray*}
where $X^+$ plays the role of a time.
The total longitudinal momentum is denoted by $P^+$,  $P^-$ is the 
light-cone energy, and $P^\perp$ the total transverse momentum. 

Let us now focus on ${\cal N}=1$ SYM(2+1). The action is
\[
S=\int d^2 x \int_0^l dx_\perp \mbox{tr}(-\frac{1}{4}F^{\mu\nu}F_{\mu\nu}+
{\rm i}{\bar\Psi}\gamma^\mu D_\mu\Psi).
\]
We decompose the spinor $\Psi$ in terms of its projections, $(\psi,\chi)$,
and use the light-cone gauge, $A^+=0$. 
The advantage of this physical gauge is that we can 
express everything in terms of the physical degrees of freedom, which are 
in the present example the fields $\psi$ and $\phi\equiv A^\perp$.
The light-cone supercharge is a two-component Majorana spinor, and is 
decomposed into the projections
\begin{eqnarray}
\label{sucharge}
Q^+&=&2^{1/4}\int dx^-\int_0^l
dx_\perp\mbox{tr}\left[\phi\partial_-\psi-\psi\partial_-
                 \phi\right],\nonumber\\
\label{sucharge-}
Q^-&=&2^{3/4}\int dx^-\int_0^l dx_\perp\mbox{tr}\left[2\partial_\perp\phi\psi
\right.\nonumber\\&&\left.
+g_{\rm YM}\left({\rm
i}[\phi,\partial_-\phi]+2\psi\psi\right)\frac{1}{\partial_-}\psi\right].
\end{eqnarray}
One can explicitly check that the supersymmetry algebra is fulfilled
\begin{equation}
\{Q^\pm,Q^\pm\}=2\sqrt{2}P^\pm,\quad \{Q^+,Q^-\}=-4P_\perp.
\label{SUSYalg}
\end{equation}
In DLCQ one discretizes the theory by 
compactifying $x^-$ on circle of period $2L=2\pi K/P^+$, 
with the harmonic resolution $K$, which is a cutoff in particle number.
In the transverse direction, one compactifies 
$x^\perp$ on circle of period $l$, with a transverse cutoff $\pm 2\pi T/l$.
The periodic boundary condition for the 
fields dictated by the supersymmetric formulation. The momentum
modes become discrete and
we can use the standard Fourier expansion for the fields $\phi$ and $\psi$.
With the usual commutation relations the supercharges become
\begin{eqnarray*}
Q^+&=&{\rm i}2^{1/4}\sum_{n^{\perp}\in {\bf Z}}\int_0^\infty\sqrt{k}
\left[b_{ij}^\dagger(k,n^\perp) a_{ij}(k,n^\perp)\right.\\
&&\left.\qquad\qquad \quad 
-a_{ij}^\dagger(k,n^\perp) b_{ij}(k,n^\perp)\right] dk,\\
Q^-&{=}&Q^-_1+g_{\rm YM}Q^-_2.
\end{eqnarray*}
The exact expressions for $Q^-_{1,2}$ are listed in Ref.~\cite{hhlp99}.
Here it suffices to note that $Q^-$ is linear in the coupling $g_{\rm YM}$.
We want to calculate the spectrum of the theory on the computer, and need
therefore finite dimensional representations of these operators. They are 
obtained by applying a truncation procedure, which defines the
(finite) Fock basis. The longitudinal momentum of a particle can 
take the values $n_{i}=0,1,2 ..., K$, and the transverse components
are $n^{\perp}_{i}=0,\pm 1,\pm 2, ... ,\pm T$.  
This symmetric truncation ensures the conservation 
of transverse parity symmetry,
which leads to exactly degenerate doublets.
The other symmetry of the theory is the so-called S-Symmetry, 
which is non-degenerate and 
is related to the orientation of the string of partons in the 
Fock state; it results in a sign change when 
the color indices of an operator are flipped. 
Additionally we have supersymmetry, so we get
a total factor of 8 savings in linear 
matrix size and the density of eigenstates will be 
much smaller, which allows for a better interpretation of results.

The program is then to construct the supercharge 
$Q^-$, apply Eq.~(\ref{SUSYalg}) to
calculate the Hamiltonian $P^-$ by squaring $Q^-$, 
and to diagonalize $P^-$ to obtain eigenvalues and -functions.
These data go into the calculation of the correlator, which we 
describe in the next section. We retrieve the continuum results
by solving the system for lager and larger $K$, reaching the continuum 
limit $K\rightarrow \infty$ eventually by extrapolation.


\section{Correlation functions}

The general expression for a correlator in light-cone formulation is
\begin{eqnarray}
&&F(x^+,x^-,x^\perp) =\nonumber \\
&&\qquad\langle 0| T^{++}(x^+,x^-,x^\perp) 
T^{++}(0,0,0)|0 \rangle.\label{expr}
\end{eqnarray}
Inspired by the simpler structure of correlators  
in the collinear limit $x^\perp = 0$ in conformal field theory, we 
apply the same limit here.
We can evaluate the expression, Eq.~(\ref{expr}), by inserting a 
complete set $| \alpha \rangle$ 
with energy eigenvalues $P^-_\alpha$ 
\begin{eqnarray*}
&&F(x^+,x^-,x^\perp=0) = \\
&&\qquad\sum_\alpha \langle 0| T^{++}(x^-,0,x^\perp=0)
| \alpha \rangle \\
&&\qquad \qquad \qquad \times
e^{-iP^-_\alpha x^+} \langle \alpha | T^{++}(0,0,0)
|0 \rangle.
\end{eqnarray*}
The momentum operator is
\begin{eqnarray*}
T^{++}(x) &=&  {\rm tr} \left[ (\partial_- \phi)^2 + {1 \over 2} \left(i
\psi \partial_- \psi  - i  (\partial_- \psi) \psi
\right)\right]\\
&=&T^{++}_B(x)+T^{++}_F(x)\,.
\end{eqnarray*}
Its boson and fermion contributions expressed in mode operators are
\begin{equation}
\frac{L}{\pi}T^{++}_B(n,m) | 0 \rangle = {\sqrt{n m} \over 2}
    a^\dagger_{ij}(n) a^\dagger_{ji} (m)| 0\rangle\label{bosons}
\end{equation}
\begin{equation}
\frac{L}{\pi}T^{++}_F(n,m) | 0 \rangle = {(n-m) \over 4}
b^\dagger_{ij}(n) b^\dagger_{ji}(m)| 0 \rangle\,.\label{fermions}
\end{equation}
The important thing to notice here is 
that only the two-particle states contribute to these operators.

\subsection{Free case}

It is instructive to consider the free case, because we know from 
conformal field theory that we should obtain a $1/r^6$ behavior.
The eigenfunctions
$|\alpha\rangle$ are now a set of free particles with mass $m$. 
The four independent sums over quantum numbers are converted to integrals
by
\[
\frac{1}{L}\sum_n\rightarrow \frac{1}{\pi}\int dk
\quad\mbox{and}\quad \frac{1}{l}\sum_{n_{\perp}}
\rightarrow \frac{1}{2\pi}\int dk_{\perp}.
\]

The evaluation of the bosonic contribution, Eq.~(\ref{bosons}), yields
\[
F(x^+,x^-,0)_B=\frac{i}{2(2\pi)^3} m^5 \left(\frac{x^+}{x^-} \right)^2
\frac{1}{x} K^2_{5/2}(mx)\,,
\]
where $x^2 =2x^-x^+$.
Analogously we can evaluate the fermion contribution, Eq.~(\ref{fermions}), 
and get
\begin{eqnarray}
F(x^+,x^-,0)_F=\frac{i}{4(2\pi)^3} m^5 \left(\frac{x^+}{x^-} \right)^2
\frac{1}{x}\nonumber\\
\qquad\times\left[ K_{7/2}(mx)K_{3/2}(mx)-K^2_{5/2}(mx) \right]\,.
\end{eqnarray}
%
\begin{figure}[ht]
\vspace*{-1.5cm}
\centerline{\psfig{file=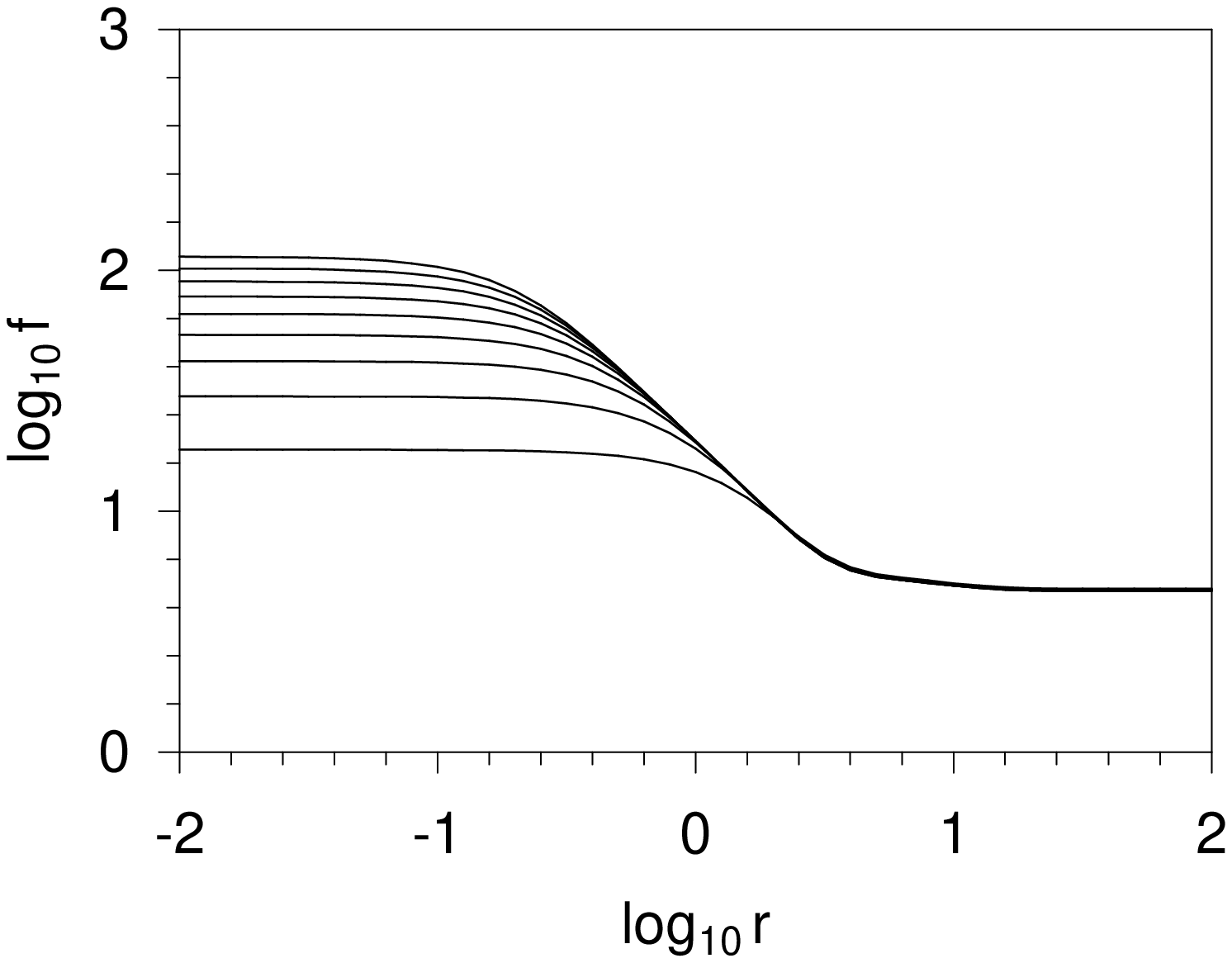,width=7cm}} 
\centerline{\psfig{file=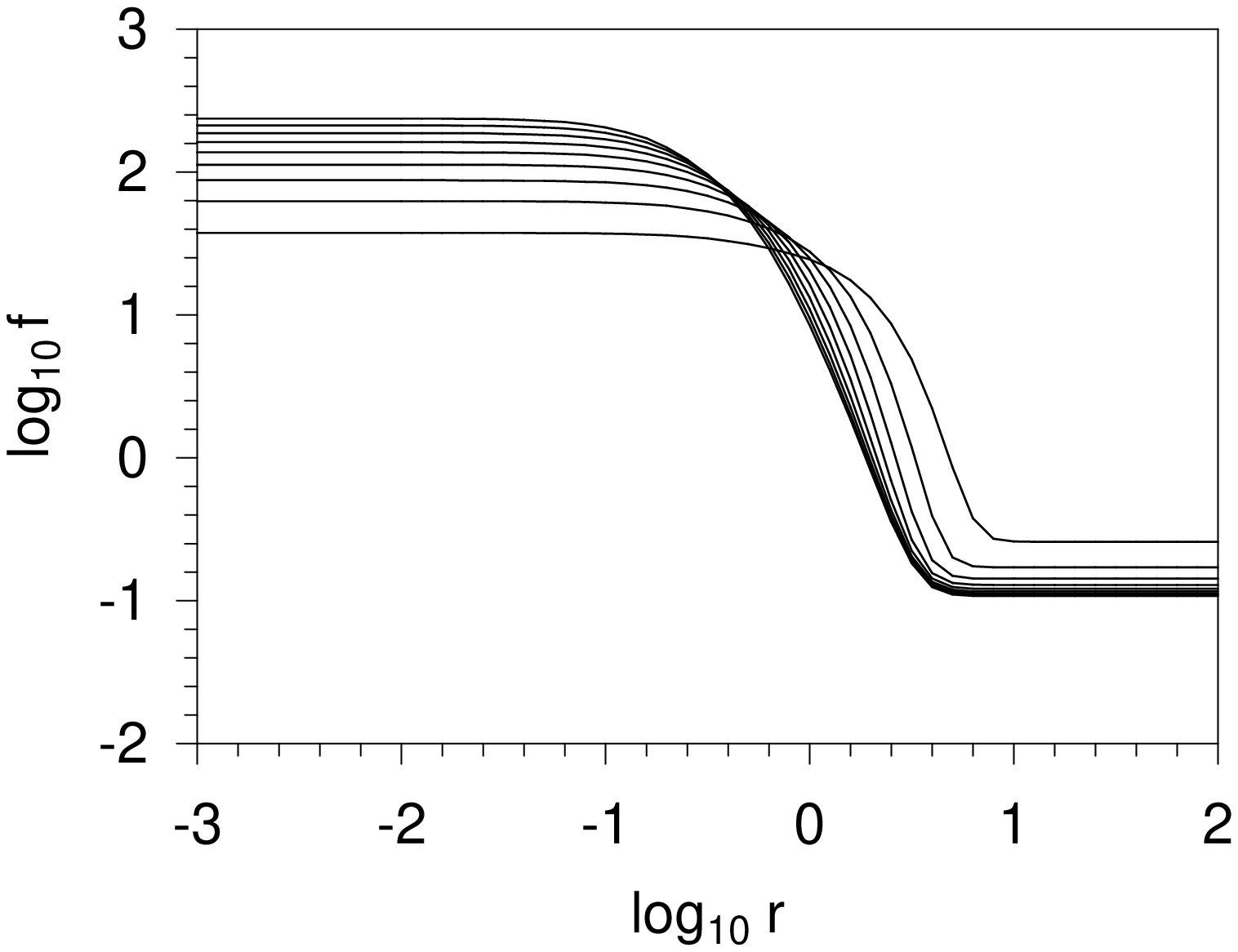,width=7cm}}
\caption{The log-log
plot of the correlation function $f\equiv r^5\langle T^{++}(x) T^{++}(0) \rangle
\left({x^- \over x^+} \right)^2 \frac{16\pi^3}{105}{{K^3 l} \over \sqrt{-i}}$
vs.\ $r$ (a) for $g{=}0.10$ at $K{=}4$ [top];
(b) for $g{=}1$ at $K=5$ [bottom]. The transverse cutoff runs from $T=1$ to 9.
\label{smallr}}
\end{figure}

We continue to Euclidean space by taking $r = \sqrt{2 x^+ x^-}$ to be real.
In the small-$r$ limit we recover the $1/r^6$ behavior
\[
\left(\frac{x^-}{x^+} \right)^2 F(x^+,x^-,0)=\frac{-3 i}{8 (2\pi)^2}
\frac{1}{r^6}\,,
\]
expected from conformal field theory.

\subsection{Full bound-state correlator}

In the full calculation we use the bound-state solutions obtained from 
SDLCQ and insert them into the expression for the correlator. 
It is useful to write  
\begin{eqnarray*} \label{eq:Fdiscrete}
&&F(x^+,x^-,0)=\sum_{n,m,s,t} \,\,\left(\frac{\pi}{2L^2 l}\right)^2\\
&&\!\!\!\!\!\!\!\!\!\!\!\!\times 
\langle 0|\frac{L}{\pi}T^{++}(n,m) e^{-iP^-_{op}x^+-iP^+x^-}
            \frac{L}{\pi} T^{++}(s,t)| 0 \rangle\,,
\end{eqnarray*}
and to use the notation
\[
\frac{1}{N_u}|u\rangle= \frac{L}{\pi}\sum_{n,m}\delta^{n+m}_{K}
\delta^{n_\perp+m_\perp}_{N_\perp}T^{++}(n,m) |0 \rangle,
\]
with a normalization factor $N_u$.
We insert the complete set of bound-states $|\alpha\rangle$
with masses $M_\alpha$. After some algebra to separate the 
dependencies on the different length scales in the problem 
we obtain
\begin{eqnarray}
F(x^+,x^-,0)&=&\frac{1}{2 \pi} \sum_{K,N_\perp,\alpha} \frac{1}{2L} 
\frac{1}{l} \left(\frac{\pi K}{L}\right)^3\\
  &&\!\!\!\!\!\!\times e^{-iP^-_\alpha x^+ -iP^+ x^-}
  \left[\frac{|\langle u|\alpha\rangle|^2}{l K^3 |N_u|^2}\right] \nonumber\,.
\end{eqnarray}
Evaluation of sums over $K$ and $N_\perp$ as integrals yields
finally
\begin{eqnarray}
&&\!\!\!\!\!\!\frac{1}{\sqrt{-i} }\left(\frac{x^-}{x^+} \right)^2 F(x^+,x^-,0)
\label{master}
=\\
\nonumber
&&\sum_\alpha 
\frac{1}{2 (2\pi)^{5/2}}\frac{M_\alpha^{9/2}}{\sqrt{r}}
K_{9/2}(M_\alpha r)
\left[\frac{|\langle u|\alpha\rangle|^2}{lK^3 |N_u|^2}\right] 
\end{eqnarray}
The term in square brackets is basically the overlap of the bound
state $|\alpha\rangle$ with the vector $|u\rangle\propto T^{++}|0\rangle$. 
It is calculated numerically and multiplied
by some function of the distance $r$ involving the Bessel function $K_{9/2}$.
It is clear from Eq.~(\ref{master}) that we need both the eigenfunctions 
$|\alpha\rangle$ and the mass eigenvalues $M_\alpha$ to evaluate this 
expression.
\begin{figure}[ht]
\centerline{\psfig{file=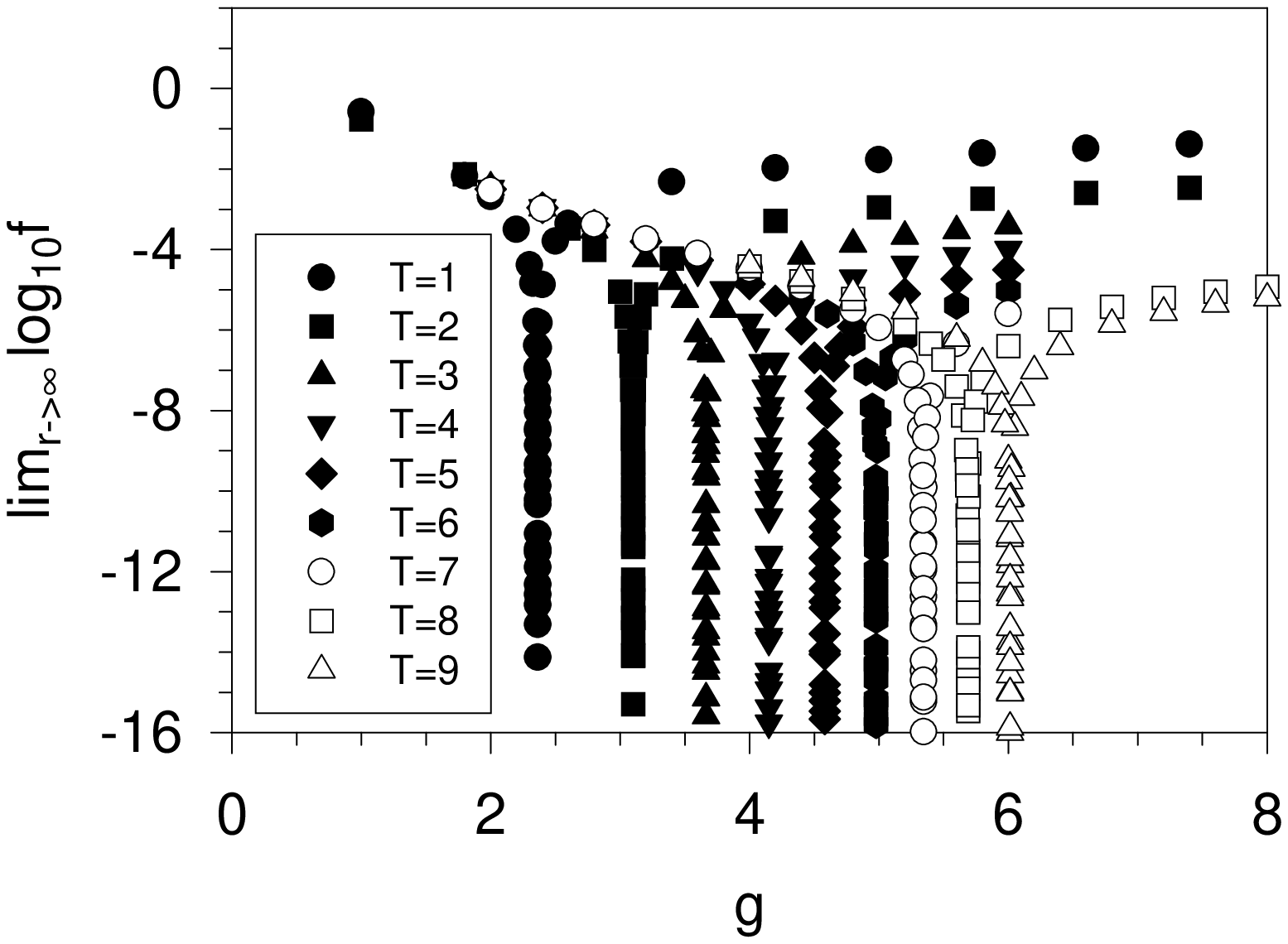,width=7cm}}
\centerline{\psfig{file=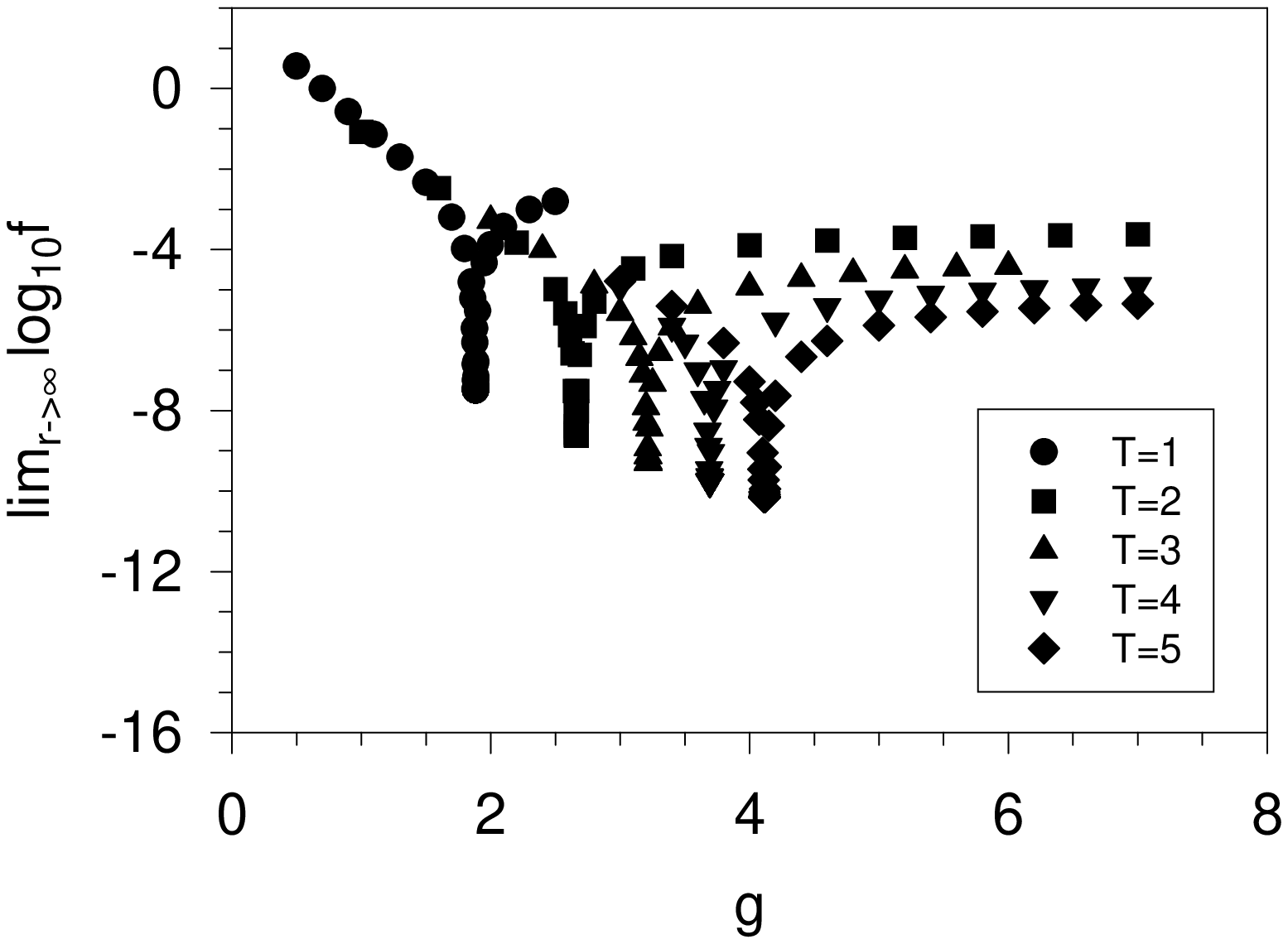,width=7cm}}
\caption{The large-$r$ limit of 
$f\equiv r^5\langle T^{++}(x) T^{++}(0) \rangle
\left({x^- \over x^+} \right)^2 
\frac{16\pi^3}{105}{{K^3 l} \over \sqrt{-i}}$ 
vs.\ $g$ for (a) $K=5$ [top]; (b) $K=6$ [bottom].
\label{k5+6large}}
\end{figure}

For free particles we have two independent sums over the transverse momenta,
and would therefore expect that the transverse dimension be 
controlled by the dimensional scale of the bound state $R_B$.
Hence, the correlation function should scale like $1/ r^4 R_B^2$.
Because of transverse boost invariance, however, 
the matrix element are independent of 
the difference of transverse momenta, and scales like $1/r^5R_B$.


\section{Numerical Results}

At small distances $r$, we
expect the correlator to behave like $1/r^6$, 
because at large energies the bound states will look like free particles.
Eq.~(\ref{master}) on the other hand tells us that 
each individual bound state behaves like $1/r^5$.  
We have to have a coherent behavior of all states 
to get the $1/r^6$ behavior, and this is a non-trivial check of our
results. Looking at Fig.~\ref{smallr}, we observe exactly this behavior.
The constant $1/r^5$ behavior of the curves at very small distances is 
caused by numerical artifacts, basically because the largest possible mass
in the system is regulated by the transverse cutoff and is not infinite. 
The slope of minus unity around $r=0$ gives rise to
the correct $1/r^6$ behavior. It consistently sets in at smaller $r$
for larger cutoffs $K$ and $T$.

At large $r$ the correlator is totally determined by the massless states.
Actually, there are two types of massless states.
The massless states at $g=0$ are reflections of all the states of 
the dimensionally reduced theory in $1+1$ \cite{alp98a}\cite{hhlp99}. 
They behave as 
$g^2 M^2_{1+1}$, and for $g \simeq 0$  there should be no
dependence of the correlator on the transverse momentum cutoff
$T$ at large $r$. This is exactly what we find in our calculations,
see Fig.~\ref{smallr}.
Secondly, we have 
states that are exactly massless for all $g$, which are the BPS states. 
They are actually massless at all resolutions, but have a
complicated dependence on the coupling $g$ through 
their wavefunctions.
In this way the correlator gives us information on wavefunctions 
of the BPS states. 

\begin{figure}[ht]
\vspace*{-0.6cm}
\centerline{\hspace*{-0.3cm}
\psfig{file=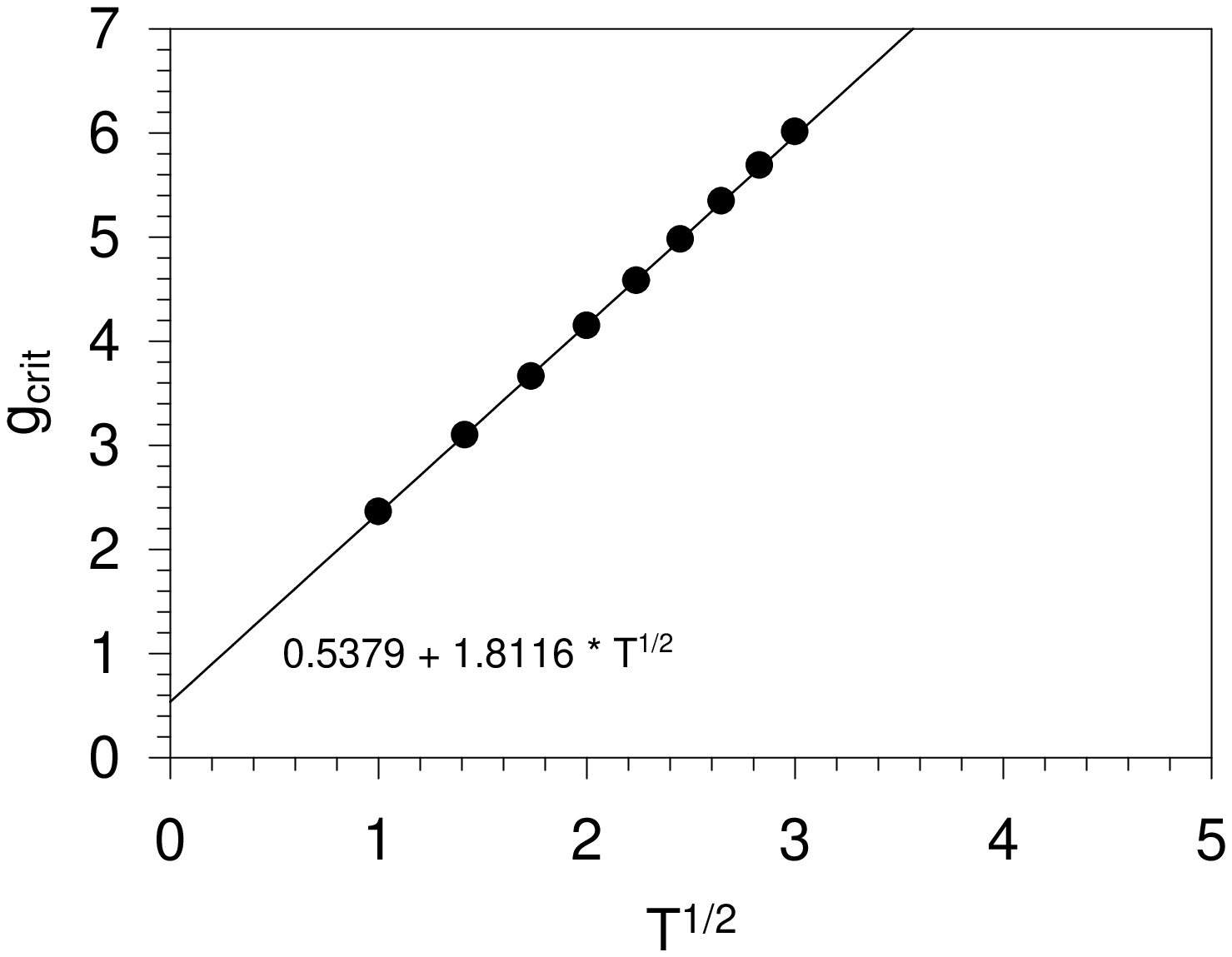,width=4cm}\hspace*{-0.3cm}
\psfig{file=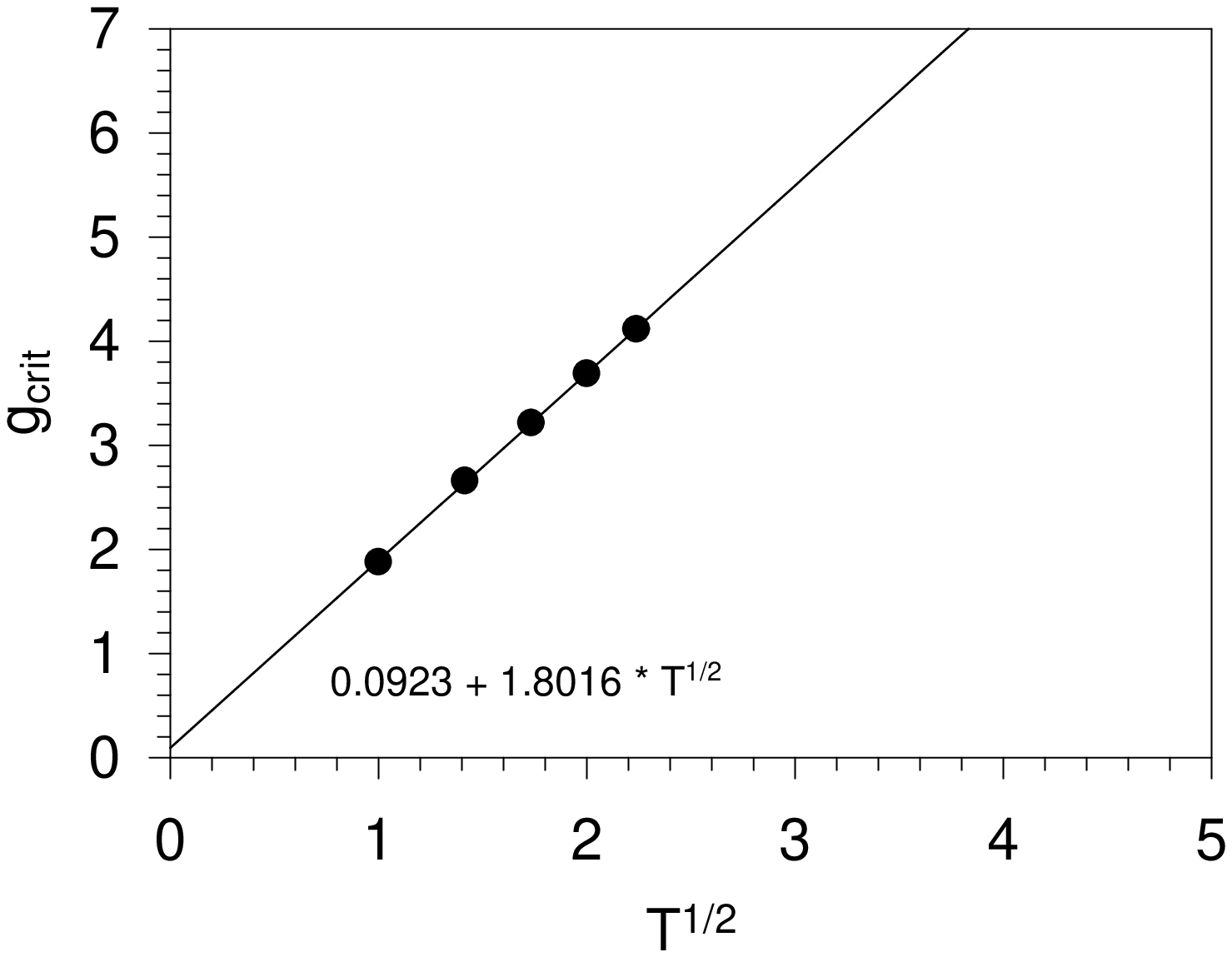,width=4cm}}
\vspace*{-0.6cm}
\caption{
Critical coupling $g_{\rm crit}$ versus $\sqrt{T}$ for (a) $K=5$ and 
(b) $K=6$.}
\label{linear}
\end{figure}

Surprisingly, we also find a coupling dependence of the 
large-$r$ limit of the correlator, see Fig.~\ref{k5+6large}.
Correlator does not change monotonically with $g$, but 
has singularity at a 'critical' coupling which is a function of $K$ and $T$.
If we plot the `critical' couplings vs.~$\sqrt{T}$ in Fig.~\ref{linear}, 
we find that the coupling is a linear function of $\sqrt{T}$ at $K=5$ and 6. 
So we would conclude that `critical' coupling goes to infinity in the
transverse continuum limit.

Unfortunately, we see no region dominated by massive bound states, 
where $r$ is large enough to see structure of the
bound states but small enough that the correlator 
is not dominated by massless states. 


\section{Conclusions}

We presented the first calculation of the correlator of the stress-energy 
tensor from first principles 
in three-dimensional ${\cal N}=1$ SYM.
We recovered the $1/r^6$ behavior of the free particle correlator, also 
known from conformal field theory.
The individual $1/r^5$ behaviors of the bound states add up to give 
this result which is a non-trivial test of our results. 
At large $r$ we found that the BPS states depend on the coupling
through their wavefunctions, altough their masses stay fixed due to their
symmetry properties.
We found a vanishing correlator at some 'critical' coupling
$g_{crit}\propto \sqrt{T}$, which seems to be a numerical artifact of the 
discrete approach.
Some remarks on the computer code seem in order.
The present code handles two million Fock states, and uses all known 
symmetries in the problem(SUSY, parity, S-symmetry), which in turn gives
a factor eight of reduction in the matrix size to be diagonalized.
Calculations in 3+1 dimensions seem therefore not completely 
out of reach.
Another challenge is the evaluation of the correlator in the 
${\cal N}=(8,8)$ version of SYM(2+1). This theory is conjectured 
to correspond to a string theory of D2-branes. 
We hope to report on progress in this direction soon, which would establish 
a further test of a version of the Maldacena conjecture. 


\end{document}